\documentclass[runningheads]{llncs}
\usepackage{graphicx}
\usepackage{float}
\usepackage{comment}
\usepackage{mdframed}
\usepackage{amsmath}
\usepackage[ruled,vlined]{algorithm2e}
\usepackage[noend]{algpseudocode}
\newtheorem{hypothesis}{Hypothesis}

\begin{document}
\title{The Bloom Clock for Causality Testing}
%
%
\author{Anshuman Misra \and Ajay D. Kshemkalyani
}
\authorrunning{Anshuman Misra and Ajay D. Kshemkalyani}
%
\institute{University of Illinois at Chicago, Chicago, IL 60607, USA\\
\email{\{amisra7,ajay\}@uic.edu}
}
\maketitle              
\begin{abstract}
Testing for causality between events in distributed executions is a fundamental problem. Vector clocks solve this problem but do not scale well. 
The probabilistic Bloom clock can determine causality between events with lower space, time, and message-space overhead than vector clock; however, predictions suffer from false positives.
We give the protocol for the Bloom clock based on Counting Bloom filters and study its properties including the probabilities of a positive outcome and a false positive.
We show the results of extensive experiments to determine how these above probabilities vary as a function of the Bloom timestamps of the two events being tested, and to determine the accuracy, precision, and false positive rate of a slice of the execution containing events in the temporal proximity of each other. Based on these experiments, we make recommendations for the setting of the Bloom clock parameters. We postulate the {\em causality spread hypothesis} from the application's perspective to indicate whether Bloom clocks will be suitable for correct predictions with high confidence. The Bloom clock design 
can serve as a viable space-, time-, and message-space-efficient alternative to vector clocks if false positives can be tolerated by an application.

\end{abstract}
\section{Introduction}
\label{intro}
\subsection{Background and Motivation}
Determining causality between pairs of events in a distributed execution is useful to many applications \cite{KS08,SM94}. This problem can be solved using vector clocks \cite{FM88,DBLP:journals/computer/Fidge91}. However, vector clocks do not scale well. 
Several works 
attempted to reduce the size of vector clocks
\cite{SK92,TA99,Meldal:1991:ELM:112600.112620,DBLP:conf/icdcn/KshemkalyaniKS18},
but they had to make some compromises in accuracy or alter the system model, and in the worst-case, were as lengthy as vector clocks. A survey of such works is included in \cite{KSV20}.

The Bloom filter, proposed in 1970, is a space-efficient probabilistic data structure that supports set membership queries \cite{B70}. The Bloom filter is widely used in computer science. Surveys of the variants of Bloom filters and their applications in networks and distributed systems are given in \cite{DBLP:journals/im/BroderM03,DBLP:journals/comsur/TarkomaRL12}. Bloom filters provide space savings, but suffer from false positives although there are no false negatives. The confidence in the prediction by a Bloom filter depends on the size of the filter ($m$), the number of hash functions used in the filter ($k$), and the number of elements added to the set ($q$). The use of the Bloom filter as a Bloom clock to determine causality between events 
was suggested \cite{DBLP:journals/corr/abs-1905-13064}, where, like Bloom filters, the Bloom clock will inherit false positives. 
The Bloom clock and its protocol based on Counting Bloom filters, which can be significantly more space-, time-, and message-space-efficient than vector clocks, 
was given in \cite{DBLP:conf/nbis/KshemkalyaniM20}. The expressions for the probabilities of a positive outcome and of a false positive as a function of the corresponding vector clocks, as well as their estimates as a function of the Bloom clocks were then formulated \cite{DBLP:conf/nbis/KshemkalyaniM20}. Properties of the Bloom clock were also studied in \cite{DBLP:conf/nbis/KshemkalyaniM20}, which then derived expressions to estimate the accuracy, precision, and the false positive rate for a slice of the execution using the events' Bloom timestamps. 

\subsection{Contributions}
In this paper, we first give the Bloom clock protocol and discuss its properties. We examine the expressions for the probability of a positive and of a false positive in detecting causality, and discuss their trends as the distance between the pair of events varies. We then show the results of our experiments to: 
\begin{enumerate}
    \item  analyze in terms of Bloom timestamps how the probability of a positive and the probability of a false positive vary as the distance between a pair of events varies; 
    \item analyze the accuracy, precision, and the false positive rate for a slice of the execution that is representative of events that are close to each other. The parameters varied are: number of processes $n$, size of Bloom clock $m$, number of hash functions $k$, probability of a timestamped event being an internal event $pr_i$, and temporal proximity between the two events being tested for causality.  
\end{enumerate}
    Based on our experiments, we 
\begin{enumerate}
    \item  analyze the nature of false positive predictions, \item make recommendations for settings of $m$ and $k$, 
    \item state conditions and analyze dependencies on the parameters (e.g., $n$, $pr_i$) under which Bloom clocks make correct predictions with high confidence (high accuracy, precision, and low false positive rate), and 
    \item generalize the above results and state a general principle (the {\em causality spread hypothesis}) based on the degree of causality in the application execution, which indicates whether Bloom clocks can make correct predictions with high confidence.
\end{enumerate}
Thus our results and recommendations can be used by an application developer to decide whether and how the application can benefit from the use of Bloom clocks.

\noindent{{\bf Roadmap:}}
Section~\ref{sysmod} gives the system model. Section~\ref{Bloomprot} details the Bloom clock protocol. Section~\ref{properties} studies properties of the Bloom clock, discusses ways to estimate the probabilities of a positive outcome and of a false positive, and predicts the trends of these probability functions as the temporal proximity between the events increases.
Section~\ref{expt} gives our experiments for the complete graph and analyzes the results.
Section~\ref{sec:star} gives our experiments for the star graph (client-server configuration) and analyzes the results.
Section~\ref{disc} summarizes the observations of the experiments and discusses the conditions under which Bloom clocks are advantageous to use. It also postulates the {\em causality spread hypothesis} and validates it.
Section~\ref{concl} concludes.

\section{System Model}
\label{sysmod}
A distributed system is modeled as an undirected graph $({\cal N}, {\cal L})$, 
where ${\cal N}$ is the set of processes and ${\cal L}$ is the set of 
links connecting them. Let $n = |{\cal N}|$.
Between any two processes, there may be at most one logical channel over which the two processes communicate asynchronously. 
A logical channel from $P_i$ to $P_j$ is formed by paths over links in ${\cal L}$. 
We do not assume FIFO logical channels. 

The execution of process $P_i$ produces a sequence of events 
$E_i = \langle e_i^0, e_i^1, e_i^2, \- \cdots\- \rangle$, where $e_i^j$ is the $j^{th}$ event at process $P_i$. 
An event at a process can be an {\em internal} event,
a {\em message send} event, or a {\em message receive} event.
Let 
$E = \bigcup_{i\in {\cal N}} \{ e \, | \, e \in E_i\}$ 
denote the set of events in a distributed execution. The causal precedence relation between events,
defined by Lamport's ``happened before'' relation \cite{L78}, and denoted as $\rightarrow$, induces an irreflexive partial order $(E, \rightarrow)$.

Mattern \cite{FM88} and Fidge \cite{DBLP:journals/computer/Fidge91}
designed the vector clock which assigns a vector $V$ to each event such that: 
$e \rightarrow f$ $\Longleftrightarrow$ $V_e < V_f$.
%
The vector clock is a fundamental tool to characterize causality in distributed executions \cite{KS08,SM94}.
Each process needs to maintain a vector $V$ of size $n$
to represent the local vector clock. Charron-Bost has shown that to capture the partial order $(E, \rightarrow)$, the size of the vector clock is the dimension of the partial order \cite{CB91}, which is bounded by the size of the system, $n$. 
Unfortunately, this does not scale well to large systems.


\section{The Bloom Clock Protocol}
\label{Bloomprot}
The Bloom clock is based on the Counting Bloom filter.
Each process $P_i$ maintains a Bloom clock $B(i)$ which is a vector $B(i)[1,\ldots,m]$ of integers, where $m$ $<n$. The Bloom clock is operated as shown in Algorithm~\ref{bc}. To try to uniquely update $B(i)$ on a tick for event $e^x_i$, $k$ random hash functions are used to hash $(i,x)$, each of which maps to one of the $m$ indices in $B(i)$. Each of the $k$ indices mapped to is incremented in $B(i)$; this probabilistically tries to make the resulting $B(i)$ unique.
As $m<n$, this gives a space, time, and message-space savings over the vector clock. We would like to point out that the scalar clock \cite{L78} can be thought of as a Bloom clock with $m = 1$ and $k = 1$.

\begin{algorithm}[t!]
    \nl Initialize $B(i) = \overline{0}$. 
    \BlankLine
    \nl (At an internal event $e^x_i$): \\
    apply $k$ hash functions to $(i,x)$ and increment the corresponding $k$ positions mapped to in $B(i)$ (local tick).
    \BlankLine
    \nl (At a send event $e^x_i$): \\
    apply $k$ hash functions to $(i,x)$ and increment the corresponding $k$ positions mapped to in $B(i)$ (local tick).
	Then $P_i$ sends the message piggybacked with $B(i)$.
	\BlankLine
    \nl (At a receive event $e^x_i$ for message piggybacked with $B'$): \\ 
    $P_i$ executes \\
	$\forall j \in [1,m], B(i)[j] = max(B(i)[j],B'[j])$ (merge);\\ 
    apply $k$ hash functions to $(i,x)$ and increment the corresponding $k$ positions mapped to in $B(i)$ (local tick). \\ 
	Then deliver the message.
    \BlankLine
	\caption{Operation of Bloom clock $B(i)$ at process $P_i$.}
    \label{bc}
\end{algorithm}

The Bloom timestamp of an event $e$ is denoted $B_e$.
Let ${\cal V}$ and ${\cal B}$ denote the sets of vector timestamps and Bloom timestamps of events. The standard vector comparison operators $<$, $\leq$, and $=$ \cite{DBLP:journals/computer/Fidge91,FM88} apply to pairs in ${\cal V}$ and in ${\cal B}$.
Thus, for example, $B_z \geq B_y$ is $\forall i \in [1,m], B_z[i]\geq B_y[i]$.
The Bloom clock mapping from $E$ to ${\cal B}$ is many-one. $({\cal B},\leq)$ is a partial order that is not isomorphic to $(E,\rightarrow)$. 

\begin{proposition}
\label{btest}
Test for $y \rightarrow z$ using Bloom clocks: if $B_z \geq B_y$ then declare $y \rightarrow z$ else declare $y \not\rightarrow z$.
\end{proposition}

\section{Properties of the Bloom Clock}
\label{properties}
We have the following cases based on the actual relationship between events $y$ and $z$, and the relationship inferred from $B_y$ and $B_z$.
\begin{enumerate}
    \item $y \rightarrow z$ and $B_z \geq B_y$: From Proposition~\ref{btest}, this results in a true positive.
    \item $y \rightarrow z$ and $B_z \not\geq B_y$: This false negative is not possible because from the rules of operation of the Bloom clock, $B_z$ must be $\geq$ $B_y$ when $y \rightarrow z$. 
    \item $y \not\rightarrow z$ and $B_z \not\geq B_y$: From Proposition~\ref{btest}, this results in a true negative. 
    \item $y \not\rightarrow z$ and $B_z \geq B_y$: From Proposition~\ref{btest}, this results in a false positive.
\end{enumerate}

Let $pr_{fp}$, $pr_{tp}$, and $pr_{tn}$ denote the probabilities of a false positive, a true positive, and a true negative, respectively. Also, let $pr_p$ denote the probability of a positive.
To evaluate these probabilities, we need $pr(y \rightarrow z)$ and $pr(B_z \geq B_y)$. As we do not have access to vector clocks, we cannot evaluate $y \rightarrow z$ as $V_y \leq V_z$. So we estimate $pr(y \rightarrow z)$ as the probability that $B_z \geq B_y$, which is the probability of a positive, $pr_p$.
So the estimate of $pr_{fp}$ is $(1-pr_p) \cdot pr_p$, from Case (4) above. However, the second term $pr_p$ can be precisely evaluated, given $B_y$ and $B_z$, as $pr_{\delta(p)}$, where 
\begin{equation}
pr_{\delta(p)} = \left\{ \begin{array}{ll}
                                     1 & \mbox{ if } B_z \geq B_y \\
                                     0 & \mbox{ otherwise}
                                     \end{array}
                            \right.
\end{equation}
So $pr_{fp} = (1-pr_p)\cdot pr_{\delta(p)}$. Also, $pr_{tp}=pr_p\cdot pr_{\delta(p)}$ from Case (1) above. Further, as a negative outcome ($B_z \not\geq B_y$) is always true from Cases (2,3) above and a negative outcome can be determined precisely, $pr_{tn} =  1-pr_{\delta(p)}$. Thus,
\begin{equation}
\begin{split}
pr_{fp} & = (1-pr_p)\cdot pr_{\delta(p)}, \\
pr_{tp} & =pr_p\cdot pr_{\delta(p)}, \\
pr_{tn} & =  1-pr_{\delta(p)}
\end{split}
\label{prdeltap}
\end{equation}
If $pr_{\delta(p)}$ were not precisely evaluated but used as a probability, we would have:
\begin{equation}
    \begin{split}
pr_{fp} & = (1-pr_p)\cdot pr_{p}, \\
pr_{tp} & =pr_p^2, \\
pr_{tn} & =  1-pr_p
\end{split}
\label{prp}
\end{equation}
We now show how to estimate $pr_p$ using Bloom timestamps $B_y$ and $B_z$. 


\begin{definition}
For a vector $X$, $X^{sum} \equiv \sum_{i=1}^{|X|} X[i]$.
\end{definition}

For a positive outcome to occur, for each increment to $B_y[i]$, there is an increment to $B_z[i]$. The number of increments to $B_y[i]$, which we denote as $c$ the {\em count threshold}, is $B_y[i]$. The probability $pr_p$ of $B_z \geq B_y$ is now formulated.
Let $b(l,q,1/m)$ denote the probability mass function of a binomial distribution having success probability $1/m$, where $l$ increments have occurred to a position in $B_z$ after applying uniformly random hash mappings $q$ times. 
\begin{equation}
    b(l,q,1/m) =
    \left( \begin{array}{c} q\\l \end{array}\right)
    (\frac{1}{m})^l(1-\frac{1}{m})^{q - l}
\label{binomialeq}
\end{equation}

Observe that the total number of trials $q = B_z^{sum}$. Then,
\begin{equation}
    b(l,B_z^{sum},1/m) =
    \left( \begin{array}{c} B_z^{sum}\\l \end{array}\right)
    (\frac{1}{m})^l(1-\frac{1}{m})^{B_z^{sum} - l}
\label{binomial}
\end{equation}
The probability that less than the count threshold $B_y[i]$ increments have occurred to $B_z[i]$ is given by:
\begin{equation}
    \sum_{l=0}^{B_{y}[i] - 1}
    b(l,B_z^{sum},1/m)
\label{lessthanthreshold}    
\end{equation}

The probability that each $i$ of the $m$ positions of $B_z$ is incremented at least $B_y[i]$ times, which gives $pr_p$, can be given by:

\begin{equation}
    pr_p(k,m,B_y,B_z) = 
    \prod_{i=1}^m 
    (1-
    \sum_{l=0}^{B_y[i] - 1}
    b(l,B_z^{sum},1/m)
    )
\label{approxbinomial}
\end{equation}

Equation
~\ref{approxbinomial} is time-consuming to evaluate for events $y$ and $z$ as the execution progresses. This is because $B^{sum}_z$ and $B_y[i]$ increase. 
A binomial distribution $b(l,q,1/m)$ can be approximated by a Poisson distribution with mean $q/m$, for large $q$ and small $1/m$. Also, the cumulative mass function of a Poisson distribution is a regularized incomplete gamma function. This provides an efficient way of evaluating Equation
~\ref{approxbinomial}.

For arbitrary event $y$ at $P_i$, to predict whether $y \rightarrow z$ where events $z$ occur at $P_j$, there are at first true negatives, then false positives, and then true positives as $z$ occurs progressively later.
As $B_z^{sum} - B_y^{sum}$ increases, we can predict the following trends from the definitions of $pr_p$ and $pr_{fp}$.
\begin{enumerate}
    \item $pr_p$, the probability of a positive, is low if $z$ is close to $y$ and this probability increases as $z$ goes further in the future of $y$. This is because, in Equation~\ref{approxbinomial}, as $B_z^{sum}$ increases with respect to $B_y^{sum}$ or rather its $m$ components, the summation (cumulative probability distribution function) decreases and hence $pr_p$ increases. 
    
    This behavior is intuitive because intuition says that as $z$ becomes more distant from $y$, the more is the likelihood that some causal relationship will get established from $y$ to $z$ either directly or transitively, by the underlying message communication pattern. 
    
    \item $pr_{fp}$, the probability of a false positive, which is the product $(1 - pr_p) \cdot pr_p$ using Equation~\ref{prp}, is lower than the two terms. It will increase, reach a maximum of 0.25, and then decrease.
    
    If Equation~\ref{prdeltap} were used, 
    then $pr_{fp} = (1-pr_p)\cdot pr_{\delta(p)}$ would be higher for a positive outcome. Once $B_z \geq B_y$ becomes true, it steps up from 0 and then as $z$ goes into the future of $y$, it decreases. 
    %
    Given a positive outcome, if $B_z \geq B_y$ and $z$ is close to $y$ ($B_z^{sum}$ is just a little greater than $B_y^{sum}$), there are two opposing influences on $pr_{fp}$: (i) it is unlikely that ``a causal relationship has been established either directly or transitively from $y$ to $z$ by the underlying message communication pattern", and thus $1-pr_p$ and $pr_{fp}$ should tend to be high; (ii) it is also unlikely that ``for each $h \in [1,m]$, $B_z[h] \geq B_y[h]$ due to Bloom clock local ticks only (and not due to causality merge for $y\rightarrow z$)", and thus $pr_{fp}$ should tend to be low. 
    As $z$ goes more distant from $y$, the likelihood of influence (i) that a causal relation has been established increases, resulting in a lower $1-pr_p$ and hence lower $pr_{fp}$. This {\em overrides} any conflicting impact of the likelihood of influence (ii), that $\forall h, B_z[h] \geq B_y[h]$ due to local ticks only and not due to causality merge for $y\rightarrow z$, increasing and thus increasing $pr_{fp}$.
    
    Based on the above reasoning, it is not apparent whether Equation~\ref{prdeltap} or~\ref{prp} is better for modeling $pr_{fp}$ behavior. However, Equation~\ref{prdeltap} uses the full range of [0,1] (as opposed to [0,0.25]), and uses an approximation only for $pr(y\rightarrow z)$ and not for $pr(B_z \geq B_y)$. 
    
    
\end{enumerate}
We remind ourselves that these probabilities depend on $B_y$, $B_z$, $k$, and $m$, and observe that they are oblivious of the communication pattern in the distributed execution.



We are also interested in calculating the accuracy, precision, and false positive rate of Bloom clocks. Accuracy ($Acc$), precision ($Prec$), recall ($Rec$), and false positive rate ($fpr$) are metrics defined over all data points, i.e, pairs of events, in the execution. Let TP, FP, TN, and FN be the number of true positives, number of false positives, number of true negatives, and the number of false negatives, respectively. Observe that FN is 0 as there are no false negatives. We have:
\begin{equation}
\label{fprbasic}
\begin{split}
Accuracy = \frac{TP+TN}{TP+TN+FP + FN},\;  & Precision = \frac{TP}{TP + FP}, \\
Recall = \frac{TP}{TP + FN},\; & fpr = \frac{FP}{FP + TN}
\end{split}
\end{equation}
Recall is always 1 with Bloom clocks. 
Given events $y$ and $z$ and their Bloom timestamps $B_y$ and $B_z$, there is not enough data to compute these metrics. So we consider the slice of the execution from $y$ to $z$ and define the metrics over the set of events in this slice.

We observe that many applications in distributed computing require testing for causality between pairs of events that are temporally close to each other. In checkpointing, causality needs to be tracked only between two consistent checkpoints. In fair mutual exclusion in which requests need to be satisfied in order of their logical timestamps, contention occurs and request timestamps need to be compared only for temporally close requests. For detecting data races in multi-threaded environments, a causality check based on vector clocks can be used; however, in practice one needs to check for data races only between events that occur in each other's temporal locality \cite{Poz19,PK21}.
In general, many applications are structured as phases and track causality only within a bounded number of adjacent phases \cite{DBLP:conf/icdcs/CouvreurFG92,DBLP:journals/ipl/Misra91}. 
Thus, in our experiments to measure accuracy, precision, and false positive rate, as well as the probability of positives and the probability of false positives, we consider an execution slice that is relatively thin. 

There is a trade-off using Bloom clocks. $m$ can be chosen less than $n$, for space, time, and message-space savings. But for acceptable precision, accuracy, and {\em fpr}, and a suitable $pr_{fp}$ distribution, an appropriate combination of values for the clock parameters $m$ and $k$ can be determined.

\section{Experiments for the Complete Graph}
\label{expt}
In the complete graph, we assume a logical channel between each pair of processes.
This experiment consists of a decentralized system of processes asynchronously passing messages to each other over shared memory. The processes are scheduled in a fair manner and are identical to each other. Even though FIFO channels are not maintained, a majority of messages arrive in order.  
The parameters of this experiment are \emph{number of processes ($n$)}, \emph{size of Bloom clock ($m$)}, \emph{internal event probability ($pr_i$)}, and \emph{number of hash functions ($k$)}. Each event can be uniquely identified with a \emph{Global Sequence Number (GSN)}. An event is modelled as an object with the following attributes:
    (i) vector timestamp, 
    (ii) Bloom timestamp,  
    (iii) GSN, 
    (iv) executing process ID,
    (v) sending process ID,
    (vi) receiving process ID,
    (vii) physical timestamp.


The main program establishes shared memory, creates $n$ processes and supplies them with parameters $pr_i$, $k$, and $m$. It then waits for all processes to complete execution and analyzes the distributed execution log.
Shared memory consists of an integer tracking GSN, a message queue containing messages (send events) yet to be received, and an execution log containing all events executed at any point of the distributed execution. All processes maintain a local queue containing messages asynchronously pulled from the shared message queue. Message receive events are executed by processing messages one at a time from the local queue with probability $(1-pr_i)/2$. Send events are executed with probability $(1-pr_i)/2$. For each send event the sending process randomly selects a receiving process from the other $n-1$ processes. Processes execute internal events with probability $pr_i$. All executed events are pushed into the global execution log. Send events are also pushed into the global message queue. 

Each process maintains its own vector clock and Bloom clock which are ticked in accordance to the vector clock and Bloom clock protocols, whenever an event is executed. The event object stores the local process's revised clocks as its vector and Bloom timestamps. In addition to this, upon executing an event, each process increments the global GSN variable by 1 and stores it in the event object. Whenever a process increments the global GSN counter, it has to acquire a lock. This is done to prevent race conditions on the GSN counter as it is stored in shared memory. Other operations that are required to be atomic and  around which locks are used include accessing the global message queue in shared memory containing messages that are waiting to be retrieved.
Each process continues to iterate and execute events until the GSN reaches $n^2$. Once all processes terminate, the main program analyzes the execution to compute precision, accuracy, and {\em fpr} of the Bloom clock protocol from the execution log. The execution log contains approximately $n^2$ events at the end of the execution. 

The main program computes causal relationships of pairs of events in the execution slice beginning with the event with GSN = $10n$ (to eliminate any startup effects) and until the last event (with GSN = $n^2)$ in steps of $100$. This means that the sample that we use to check for causality predictions consists of a series of events where two closest events have a difference of $100$ in GSN. Further, the number of pairs of events for which we tested for causality was approximately $n^4/10^4$.
The main program compares causality predictions of the Bloom timestamps of events with predictions of vector timestamps and classifies the Bloom clock predictions as true positives, false positives and true negatives. The precision, accuracy, and {\em fpr} are computed over this execution slice. We intentionally chose an execution slice with $n$ events per process because in practice, causality tests are applied to pairs of events in the temporal proximity of each other. Had we chosen a larger execution slice, we expect the metrics would have improved. 

Finally, in this section and the next on experiments with the star configuration, each reading reported is the average of at least 3 runs of each setting of the parameters indicated.
Also, in Sections~\ref{sec:pri} to~\ref{sec:m}, where indicated, each reported reading is also averaged over multiple settings of $m$ and/or $k$ for simplicity of presentation of results; the impact of varying each individual parameter is clear when the results of all experiments are considered.

\subsection{Number of Processes}

\begin{figure}[h!]
  \centering
  
    \includegraphics[width=\linewidth]{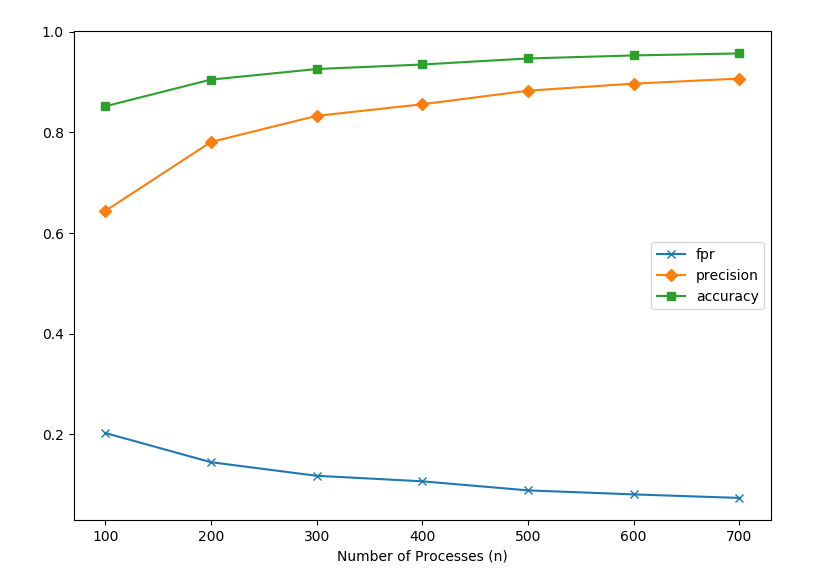}
    \caption{A plot of metrics vs. number of processes for decentralized execution}
  
  \label{fig:n_plot}
\end{figure}

We ran the decentralized experiment for $n = 100$ to $n = 700$ in increments of $100$ to ascertain scalability of Bloom clocks. Parameters were fixed to maintain uniformity of results with $pr_i = 0$, $k = 2$, and $m = 0.1*n$. The results are compiled in Table~\ref{table:n_trend}. A visual representation of the trend can be seen in Figure \ref{fig:n_plot}. We see that as $n$ increases Bloom clock performance improves considerably.  
Accuracy increases from $85.2\%$ for $n=100$ to $95.7\%$ for $n=700$ and 
the {\em fpr} drops from $20.3\%$ for $n=100$ to $7.4\%$ for $n=700$. 
Since Bloom clocks are not prone to false negatives, a critical method of measuring performance is to calculate the ratio of positive predictions that are correct to overall positive predictions. 
Precision measures exactly that. 
We observe that precision increases from $64.4\%$ for $n=100$ to $90.7\%$ for $n=700$. 
Overall from Table~\ref{table:n_trend}, we conclude that Bloom clocks are highly scalable.

\begin{table}[H]
\caption{Variation of metrics with $n$} 
\centering 
\begin{tabular}{c c c c} 
\hline 
$n$ & Precision & Accuracy & {\em fpr} \\ [0.5ex] 
\hline 
100 & 0.644 & 0.852 & 0.203 \\ 
200 & 0.781 & 0.905 & 0.145 \\
300 & 0.833 & 0.926 & 0.118 \\
400 & 0.856 & 0.935 & 0.107 \\
500 & 0.883 & 0.947 & 0.089 \\
600 & 0.897 & 0.953 & 0.081 \\
700 & 0.907 & 0.957 & 0.074 \\
\hline

\end{tabular}
\label{table:n_trend} 
\end{table}

\subsection{Internal Event Probability}
\label{sec:pri}
We ran the decentralized experiment for fixed $n=200$ and averaged metrics over $m=0.1*n, 0.2*n, 0.3*n$ and $k=2, 3, 4$ for individual values of $pr_i$ in order to observe the variation of metrics with $pr_i$. The results are shown in Table~\ref{table:p_trend}. We observed that by introducing more relevant (and therefore timestamped) internal events in the decentralized execution, the performance of Bloom clocks deteriorates significantly. So with an increase in send events and thus message-passing, i.e., a relative decrease in the number of relevant timestamped internal events, more causal relationships get established among events across processes, which get captured through the merging of Bloom clocks at receive events. This results in a higher fraction of the number of pairs of events being related by causality and a smaller fraction of the number of pairs of events being concurrent. Bloom clocks performed best at $pr_i = 0$. 
We generalize this observation as the {\em causality spread hypothesis} later in Section~\ref{sec:causalityspread}.

\begin{table}[h]
\caption{Variation of metrics with $pr_i$} 
\centering 
\begin{tabular}{c c c c} 
\hline 
$pr_i$ & Precision & Accuracy & {\em fpr} \\ [0.5ex] 
\hline 
0 & 0.807 & 0.918 & 0.125 \\
0.90 & 0.609 & 0.847 & 0.201 \\
0.95 & 0.311 & 0.760 & 0.269 \\
1 & 0.101 & 0.773 & 0.232 \\
\hline

\end{tabular}
\label{table:p_trend} 
\end{table}

The practical implication of setting $pr_i=0$ is that most of the relevant events at which clocks tick are send and receive events, and only a few internal events (of interest to the application) cause the clocks to tick.
In contrast, with a high value of $pr_i$ (such as 0.9 at which $90\%$ of events at which clocks tick are internal events), accuracy and precision drop significantly, and {\em fpr} increases significantly. Thus, Bloom clocks are practical only when the percentage of relevant events (where clock ticks) that are internal events is small.


\subsection{Number of Hash Functions}
\label{sec:k}
We ran the decentralized experiment for fixed $n = 200$ and fixed $pr_i = 0$ and averaged metrics over $m = 0.1*n, 0.2*n, 0.3*n$  for individual values of $k$ to check the variation of Bloom clock performance with respect to $k$. The results are shown in Table \ref{table:k_trend}. We observe that the effect of changing the number of hash functions does not have a quantifiable effect on Bloom clock performance. 

\begin{table}[h]
\caption{Variation of metrics with $k$} 
\centering 
\begin{tabular}{c c c c} 
\hline 
$k$ & Precision & Accuracy & {\em fpr} \\ [0.5ex] 
\hline 
2 & 0.804 & 0.917 & 0.126 \\
3 & 0.809 & 0.919 & 0.124 \\
4 & 0.808 & 0.919 & 0.124 \\
\hline


\end{tabular}
\label{table:k_trend} 
\end{table}

\subsection{Size of Bloom Clock}
\label{sec:m}
We ran the decentralized experiment for fixed $n = 200$ and fixed $pr_i = 0$ and averaged metrics over $k=2, 3, 4$ for individual values of $m$ to check the variation of Bloom clock performance with respect to $m$. The results are shown in Table~\ref{table:m_trend}. As expected, Bloom clock performance improves, but by up to $4.3\%$ points, as $m$ increases from $0.1*n$ to $0.3*n$. The improvement seems intuitive because with a larger number of indices the probability of hash function outputs mapping to the same indices reduces, due to which there is a lower probability of false positives.

\begin{table}[h]
\caption{Variation of metrics with $m$} 
\centering 
\begin{tabular}{c c c c} 
\hline 
$m$ & Precision & Accuracy & {\em fpr} \\ [0.5ex] 
\hline 
$0.1*n$ & 0.784 & 0.906 & 0.143 \\
$0.2*n$ & 0.811 & 0.920 & 0.122 \\
$0.3*n$ & 0.827 & 0.929 & 0.109 \\
\hline


\end{tabular}
\label{table:m_trend} 
\end{table}

In addition, we ran the experiment with scalar clock ($m$ = 1 and $k$ = 1) instead of Bloom clock, in order to investigate improvement in metrics for Bloom clock over scalar clock. We compared Bloom clock of size $m = 0.1*n$ and $k = 2$ to scalar clock at various values of $n$ for ${pr}_i = 0$. The results are presented in Table \ref{table:scalar_clock}. We observe significant performance improvements over scalar clock by utilizing Bloom clock at all values of $n$ -- precision was 0.06 to 0.11, accuracy was 0.07 to 0.09, and {\em fpr} was 0.10 to 0.12 better.

\begin{table}[h]
\caption{Bloom clock vs. scalar clock} 
\centering 
\begin{tabular}{c | c c c | c c c} 
\hline 
&

\multicolumn{3}{c|}{Bloom Clock}
&
\multicolumn{3}{c}{Scalar Clock} \\
\hline
$n$ & Precision & Accuracy & {\em fpr} &  Precision & Accuracy & {\em fpr}  \\ [0.5ex] 
\hline 
$50$ & 0.492 & 0.788 & 0.266 & 0.434 & 0.713 & 0.368 \\
$100$ & 0.644 & 0.852 & 0.203 & 0.542 & 0.769 & 0.318 \\
$200$ & 0.781 & 0.905 & 0.145 & 0.672 & 0.835 & 0.248 \\

\hline 
\end{tabular}
\label{table:scalar_clock} 
\end{table}

\subsection{Plots for $pr_p$ and $pr_{fp}$}

\begin{figure}[H]
  \centering
  
    \includegraphics[width=\linewidth]{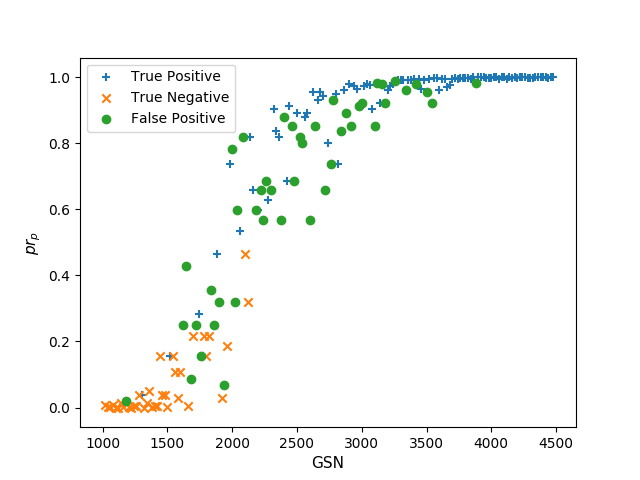}
    \includegraphics[width=\linewidth]{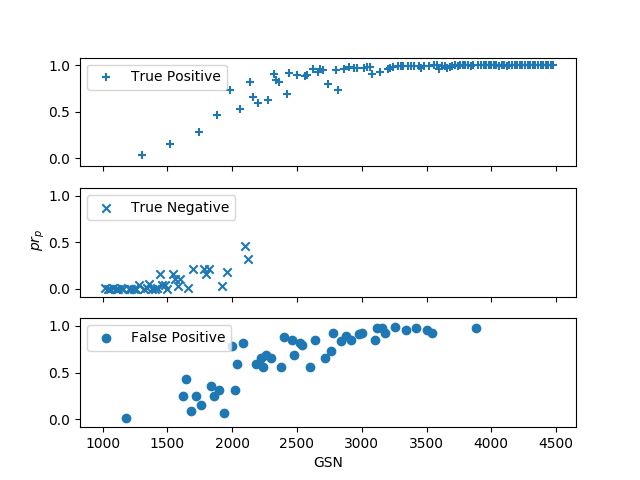}
    \caption{$pr_p$ vs. GSN, showing combined view and split view}
  
  \label{fig:prp2}
\end{figure}

\begin{figure}[H]
  \centering
  
    \includegraphics[width=\linewidth]{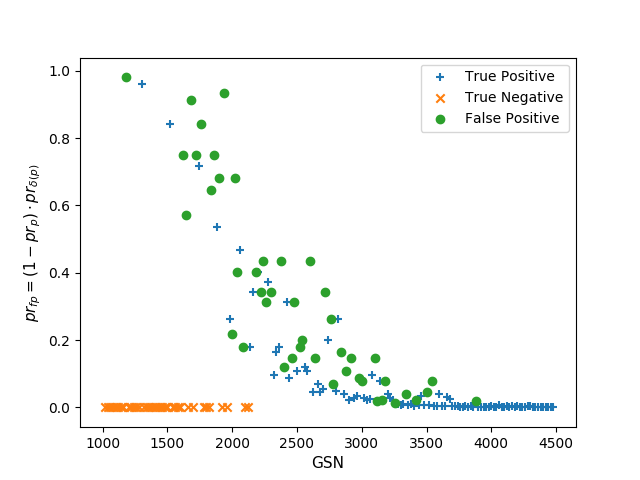}
    \includegraphics[width=\linewidth]{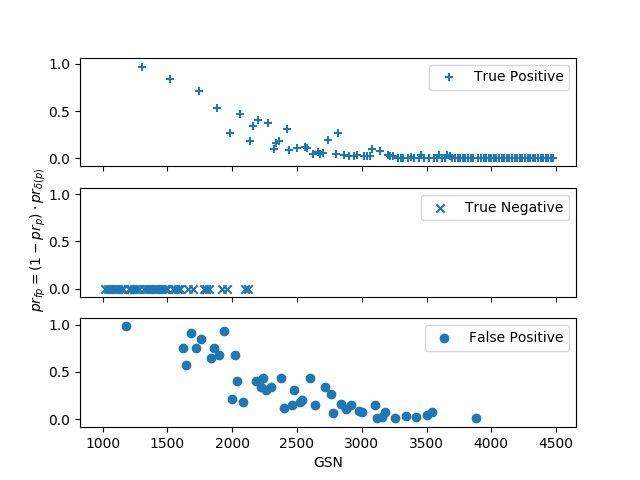}
    \caption{$pr_{fp} = (1-pr_p) \cdot pr_{\delta(p)}$ using Equation~\ref{prdeltap} vs. GSN, showing combined view and split view}
  
  \label{fig:prfp2}
\end{figure}

\begin{figure}[H]
  \centering
  
    \includegraphics[width=\linewidth]{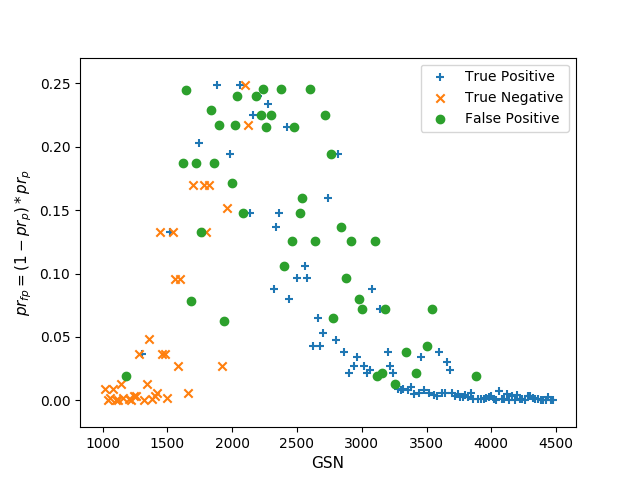}
    \includegraphics[width=\linewidth]{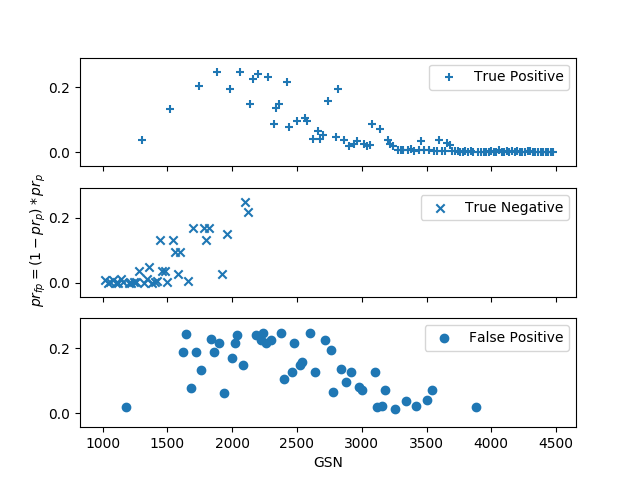}
    \caption{$pr_{fp} = (1-pr_p) \cdot pr_{p}$ using Equation~\ref{prp} vs. GSN, showing combined view and split view}
  
  \label{fig:prfp4}
\end{figure}

We ran the decentralized experiment for fixed parameters $n=100$, $pr_i=0$, $k=2$ and $m=0.1*n$ to obtain plots for $pr_p$, and $pr_{fp}$ computed using Equations~\ref{prdeltap} and~\ref{prp}. These plots demonstrate the behavior of Bloom clocks throughout an execution as the temporal proximity between events $y$ and $z$ varies, using just the Bloom timestamps of the two events being compared for causality. 
For these plots we fix event $y$ with $GSN = 10*n$, which is 1000, to allow for any startup transient effects, and compare its Bloom timestamp with all events $z$ with $GSN = 10*n + 1$ to $GSN = 4500 \,(\sim n^2/2)$. This slice of the execution is adequate to capture all the trends. The x-axis of Figures~\ref{fig:prp2} to~\ref{fig:prfp4} is the GSN of $z$ and the y-axis is the probability being plotted. 

Figure~\ref{fig:prp2} shows a plot of $pr_p$ as a function of GSN. We observe that as GSN increases, the probability of a positive prediction increases and flattens to around $1$ between GSN = 3500 and GSN = 4000. This is because as the distance between two events increases, there is a higher probability of a causal relationship being established either directly or transitively. 
The split view of $pr_p$ vs. GSN allows us to observe that most false positives occur in the middle of the distribution while all true negatives occur within the first half of the execution. This is due to the fact that initially the probability of a true negative is very high because the probability of a causal relationship being established is lower.

Figure~\ref{fig:prfp2} 
shows plots for $pr_{fp} = (1-pr_p)\cdot pr_{\delta(p)}$ (Equation~\ref{prdeltap}) vs. GSN. We observe that Bloom clocks correctly predict the probability of false positive being 0 for all true negatives in the execution. Most of the false positives are distributed in the middle of the execution slice; the $pr_{fp}$ jumps from $0$ to large values once false positives start occurring and then gradually decreases as GSN increases. The (few) false positives that occur towards the end of the execution slice are not captured correctly with low values of $pr_{fp}$. The probability of false positive for a majority of true positives is below $0.25$; however, for the initial few true positives, the $pr_{fp}$ is inaccurately evaluated as being high. This probability $pr_{fp}$ (for the true positives) rapidly decreases to 0 as GSN increases.

Figure~\ref{fig:prfp4} 
shows plots for $pr_{fp} = (1-pr_p)*pr_p$ (Equation~\ref{prp}) vs. GSN. As expected, $pr_{fp}$ has values below $0.05$ for most true negatives and true positives and reaches a maximum value of $0.25$ in the middle of the execution where most of the false positives reside. Thus, the $pr_{fp}$ is inaccurately evaluated as being low for the false positives in the middle of the execution slice.

Thus, Figures~\ref{fig:prp2} to~\ref{fig:prfp4} confirm the theoretical predictions made in Section~\ref{properties}. Equation~\ref{prdeltap} uses a range of [0,1] for $pr_{fp}$, gives a high $pr_{fp}$ to the initial few true positives, and does not seem to capture the two conflicting influences on $pr_{fp}$ described in Section~\ref{properties} when the GSN of $z$ is just a little greater than the GSN of $y$. Equation~\ref{prp} uses a range of only [0,0.25] and inaccurately gives a low $pr_{fp}$ for the false positives in the middle of the execution slice.

\section{Experiments for the Star Graph}
\label{sec:star}
We set up an experiment with a client-server architecture to investigate how faithfully the Bloom clock determines causality. Client processes connect to a multi-threaded server accepting TCP connections. Each server thread connects to a single client. 
The internal event probability, $pr_i$ was set to $0$. 
All message sends were synchronous and blocking and receives were blocking. 
Each client consisted of a process and had its own vector clock and Bloom clock. The server had a single vector clock and a single Bloom clock shared across all threads. 

The server threads used a single lock to make sure that there were no race conditions on the vector clock and the Bloom clock while executing events. 
We did not use locks at the client end because GSN was not maintained. Further, not using locking mechanisms allowed interleaving of client processes.

Each client sent $n$ messages to the server and received $n$ corresponding messages from the server. This resulted in overall $O(n^2)$ events in the execution. Post execution, each $100^{th}$ event was taken from the execution log containing all the events from the execution to create a sample of events to be compared for causality. Each event $y$ was compared to each other event $z$ to determine if Bloom clock correctly classified whether $y \to z$ or $y \not\to z$. The correctness of the Bloom clock prediction was ascertained by comparing it with the prediction from vector clock. The results for the client-server experiment for $k=2$ are shown in Table~\ref{table:client_server}.

As can be seen from the results, Bloom clock performs quite well with high values of precision and low {\em fpr}. The first four rows are for $m = 0.1*n$ and the last four rows are for $m = 0.05*n$. We observed that for a small Bloom clock of size $m=3$ for $n=50$, the accuracy is high at $100\%$ (There was one false positive, but rounding off to three decimal places results in the stated accuracy value). The difference in precision, accuracy, and {\em fpr} for smaller Bloom clocks as compared to larger Bloom clocks is not significant, therefore it is safe to say that for this configuration, smaller Bloom clocks perform well. The reason for strong performance of Bloom clock is that there are a lot of merge events with a centralized process, and the inherent message pattern at the server resulted in automatic and widespread distribution/broadcasting of information contained in individual Bloom clocks among all client processes. The server is always up to date with a client's Bloom clock after it executes a receive event corresponding to a message send event from the client. We generalize the reasoning behind the good performance of the Bloom clock for the client-server configuration by postulating the {\em causality spread hypothesis} in Section~\ref{sec:causalityspread}.

\begin{table}[h]
\caption{Results for client-server experiment with $k = 2$} 
\centering 
\begin{tabular}{c c c c c} 
\hline 
$n$ & $m$ & Precision & Accuracy & {\em fpr} \\ [0.5ex] 
\hline 
50 & 5 & 0.985 & 0.992  & 0.015 \\
100 & 10 &  0.990  & 0.995 & 0.010 \\
125 & 13 & 0.991 & 0.996 & 0.009 \\
150 & 15 & 0.995 & 0.997 & 0.005 \\
50 & 3 & 100 & 100 & 0\\
100 & 5 & 0.996  & 0.998 & 0.004 \\
125 & 7 & 0.997 & 0.998  & 0.003 \\
150 & 8 & 0.997 & 0.998 &  0.003 \\
\hline 
\end{tabular}
\label{table:client_server} 
\end{table}

\section{Observations and Discussion}
\label{disc}
\subsection{Summary of Results}
\label{analysis}
The results of the experiments are summarized as follows.
\begin{enumerate}
    \item In predicting the causality between events $y$ and $z$ using their Bloom timestamps, we observe the following.
    \begin{enumerate}
        \item The probability of a positive $pr_p$ increases relatively quickly from 0 to 1 as $z$ occurs after but in the temporal vicinity of $y$.
        \item The probability of a false positive $pr_{fp}$ is 0 or close to 0 except when $z$ occurs later than but in the temporal vicinity of event $y$. As $z$ occurs later at a process, the probability spikes up from 0 to a high value but soon comes down to 0 as the occurrence of $z$ get temporally separated from the occurrence of $y$. Some true positives have a non-zero value of $pr_{fp}$. 
    \end{enumerate}
    \item As the number of processes $n$ increases, the Bloom clock performance improves significantly -- the accuracy and precision increase, and the {\em fpr} decreases.
    \item When the number of internal events at which the clock ticks is low relative to the number of send  events, precision, accuracy, and {\em fpr} all improve significantly. Thus, with relatively more send events, performance of Bloom clock improves. With more send events, causality between more pairs of events is established. 
    On the other hand, if the number of internal events being timestamped is high with respect to the number of send events, Bloom clocks do not perform well. 
    \item The number of hash functions $k$ used in the Bloom clock protocol does not impact much the precision, accuracy, and the {\em fpr}. Hence, it is advantageous to use a small number (such as 2 or 3) of hash functions.
    \item The precision, accuracy, and the {\em fpr} improved by a few percentage points as the size of the Bloom Clock $m$ was increased from $0.1*n$ to $0.3*n$. The impact is noticeable but not much. Hence, this suggests that small-sized Bloom Clocks can be used to gain significant space, time, and message-space savings over vector clocks. As a baseline for comparison, we also measured the precision, accuracy, and {\em fpr} for Lamport's scalar clocks. The scalar clocks performed noticeably worse.
    \item For the client-server configuration, Bloom clocks performed exceedingly well.
\end{enumerate}
Bloom clocks are seen to provide a viable space-, time-, and message-space-efficient alternative to vector clocks when some false positives can be tolerated. Bloom clock metrics improve as the number of processes increases. Bloom clock sizes can be $10\%$ or even lower of the number of processes, and can handle churn transparently when processes join and leave the system. The probability of a false positive is high only when the two events occur temporally very close to each other. However, Bloom clocks do not perform well when the fraction of timestamped events that are internal events is not very low. In the next section, we generalize this behavior using the {\em causality spread hypothesis}.

\subsection{Causality Spread}
\label{sec:causalityspread}
After conducting experiments to track causality using the Bloom clock for multiple architectures and varying parameters, we develop a hypothesis to help system engineers and software developers figure out whether the Bloom clock is a good fit for a given application. This hypothesis is stated only from the application's perspective.
We hypothesize that with an increase in spread of causality in an execution, i.e., with a larger proportion of events related by causal relationships, Bloom clock performance (i.e., confidence in its predictions) increases. 
We define and compute the \emph{causality spread}, $\alpha$, as the ratio of the number of ordered pairs of events that are causally related, that is, \emph{total positives}, to the sum of all ordered pairs of events compared for each execution.
The set of events that we include in the computation of causality spread are the {\em relevant events} for the application.

\begin{definition}[Causality spread $\alpha$]
\label{def:cs}
\begin{equation}
  \begin{split}
\text{Causality spread } \alpha & = \frac{\text{Total Positives}}{\text{\# All pairs of events}}\\
 & = \frac{\text{Total Positives}}{\text{Total Positives + Total Negatives}}\\
 & = \frac{\text{TP + FN}}{\text{TP + FN + FP + TN}} = \frac{\text{TP}}{\text{TP + FN + FP + TN}}
\end{split}
  \label{causal_spread}
\end{equation}
\end{definition}

\begin{hypothesis}[Causality spread hypothesis]
\label{cshypothesis}
 The confidence in the predictions of the Bloom clock as measured by precision, accuracy, and {\em fpr} increases as the causality spread $\alpha$ of the application's set of relevant events increases.
\end{hypothesis}

A higher $\alpha$ signifies more (fraction of) event pairs being related by causality, which are correctly classified as true positives, thereby increasing TP (say, by $a$), decreasing FP, decreasing TN, and decreasing FP + TN (by $a$). Theoretically, we expect precision and accuracy will improve (as per some non-linear functions), while the impact on {\em fpr} depends on the factors by which its numerator FP and its denominator FP + TN change. 

\begin{figure}[t]
  \centering
  
    \includegraphics[width=\linewidth]{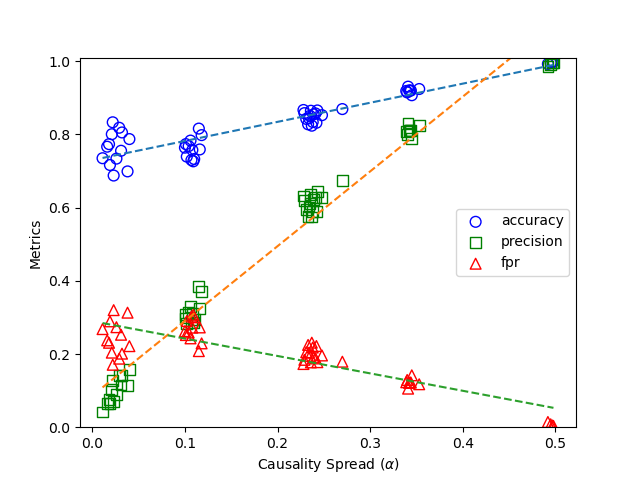}
    \caption{A plot of metrics vs. {\em causality spread}}
  
  \label{fig:cs}
\end{figure}

This hypothesis is corroborated by our previously stated observation that increased message passing results in superior Bloom clock predictions.
In order to quantify this hypothesis, we took a sample of executions from both the decentralized experiment and the client-server experiment and computed $\alpha$.
We observed that precision and accuracy increase and \emph{fpr} decreases as $\alpha$ increases, for $0 < \alpha < 0.5$, empirically confirming our hypothesis. A graph showing the increase in metrics as a function of causality spread is shown in Figure~\ref{fig:cs}. 

An important note about causality spread is that it will range between $0$ and $0.5$ in our experiment because we check for causality between all pairs of events. An extreme case where $\alpha = 0$ would be each process executing only one event. Another extreme case where $\alpha = 0.5$ would be a linear chain of events. 
In the client-server experiment, $\alpha$ is near 0.5 due to the nature of transmission of causal relationships because of the server behavior. In the complete graph configuration with a high $pr_i$, the large number of timestamped internal events in the set of relevant events  significantly increases the number of pairs of concurrent events and hence decreases $\alpha$ considerably, resulting in poor prediction by Bloom clocks.

We performed an experiment for multicast/broadcast messages to check if it conforms to our causality spread hypothesis. In the broadcast experiment, each process broadcasts a message to all $n - 1$ processes and waits to receive $n - 1$ broadcast messages from the other processes. Here, causality does not spread much because there is only one message send event followed by many receive events for each process. Here the receive events act as internal events that are timestamped (akin to high $pr_i$), and in effect there are many pairs of events that are concurrent and hence not related by causality, thereby resulting in a low $\alpha$. In the experiment, $\alpha = 0.005$, precision = $0.014$, accuracy = $0.661$, and fpr = $0.341$. The poor performance of Bloom clock in this experiment can be attributed to a low $\alpha$ as per the hypothesis. 

\section{Conclusions}
\label{concl}
Detecting the causality relationship between a pair of events in a distributed execution is a fundamental problem. To address this problem in a scalable way, this paper gave the formal Bloom clock protocol, and derived the expression for the probability of false positives, given two events' Bloom timestamps.  
We ran experiments to calculate the accuracy, precision, and  {\em fpr} for a slice of the execution. We also ran experiments to calculate the probability of a false positive prediction based on the Bloom timestamps of two events.
Based on the experiments, we made suggestions for the number of hash functions and size of Bloom clocks and identified conditions under which it is advantageous to use Bloom clocks over vector clocks. The findings are summarized as follows.
\begin{enumerate}
    \item Bloom clocks can perform well for small size $m$ and small number of hash functions $k$.
    \item Bloom clocks perform well when the number of internal events considered is low compared to the number of send events (low $pr_i$). 
    \item Bloom clocks perform increasingly better as the system size $n$ increases. 
    \item We also postulated the {\em causality spread hypothesis} from the application's perspective to determine whether Bloom clocks would give good performance (precision, accuracy, and {\em fpr}) for the application, and validated it through experiments. A high $\alpha$ indicates good performance.
\end{enumerate}

Thus, Bloom clocks are seen to provide a viable space-, time-, and message-space-efficient alternative to vector clocks for the class of applications which meet the properties summarized above, when some false positives can be tolerated.
It would be interesting to study the applicability of Bloom clocks to some practical applications.
\bibliographystyle{splncs04}
\bibliography{references}
%




\end{document}